\documentclass{article}%
\usepackage{amsfonts}
\usepackage{amsmath}%
\usepackage{amssymb}%
\usepackage{graphicx}
\providecommand{\U}[1]{\protect\rule{.1in}{.1in}}
\newtheorem{theorem}{Theorem}

\newtheorem{lemma}[theorem]{Lemma}

\begin{document}

\title{Test of Purity by LOCC}
\author{Keiji Matsumoto\\National Institute of Informatics, Tokyo, Japan, and JST, Tokyo, Japan}
\maketitle

\begin{abstract}
Given $n$-copies of unknown bipartite ( possiblly mixed ) state, our task is
to test whether the state is a pure state of not. Allowed to use the global
operations, optimal one-sided error test is the projection onto the symmetric
subspace, obviously. Is it possible to approximate the globally optimal
measurement by LOCC when $n$ is large?

\end{abstract}

\section{Introduction}

Given $n$-copies of unknown bipartite ( possiblly mixed ) state, our task is
to test whether the state is a pure state of not. Allowed to use the global
operations, optimal one-sided error test is the projection onto the symmetric
subspace, obviously. Is it possible to approximate the globally optimal
measurement by LOCC when $n$ is large?

\section{A standard form of an ensemble of identical bipartite pure states}

Suppose we are given $n$-copies of unknown pure bipartite state $\left\vert
\phi\right\rangle \in\mathcal{H}_{A}\otimes\mathcal{H}_{B}$, which is unknown.
Here we assume $\mathcal{H}_{A}\simeq\mathcal{H}_{B}\simeq\mathcal{H}$ and
$\dim\mathcal{H}=d$. It is known that $\left\vert \phi\right\rangle ^{\otimes
n}$ has the standard form defined as follows. 

Note $|\phi\rangle^{\otimes n}$ is invariant by the reordering of copies, or
the action of the permutation $\sigma$ in the set $\{1,\ldots n\}$ such that
\begin{equation}
\bigotimes_{i=1}^{n}|h_{i,A}\rangle|h_{i,B}\rangle\mapsto\bigotimes_{i=1}%
^{n}|h_{\sigma^{-1}(i),A}\rangle|h_{\sigma^{-1}(i),B}\rangle,\label{sym}%
\end{equation}
where $|h_{i,A}\rangle\in\mathcal{H}_{A}$ and $|h_{i,B}\rangle\in
\mathcal{H}_{B}\;$. Action of the symmetric group occurs a decomposition of
the tensored space $\mathcal{H}^{\otimes n}$ ~\cite{GW},
\[
\mathcal{H}^{\otimes n}=\bigoplus_{\lambda}\mathcal{W}_{\lambda}%
,\;\mathcal{W}_{\lambda}:=\mathcal{U}_{\lambda}\otimes\mathcal{V}_{\lambda},
\]
where $\mathcal{U}_{\lambda}$ and $\mathcal{V}_{\lambda}$ is an irreducible
space of the tensor representation of $\mathrm{SU}(d)$, and the representation
(\ref{sym}) of the symmetric group, respectively, and
\[
\lambda=(\lambda_{1},\ldots,\lambda_{d}),\quad\;\lambda_{i}\geq\lambda
_{i+1}\geq0,\,\sum_{i=1}^{d}\lambda_{i}=n
\]
is called \textit{Young index}, which $\mathcal{U}_{\lambda}$ and
$\mathcal{V}_{\lambda}$ uniquely corresponds to. We denote by $\mathcal{U}%
_{\lambda,A}$, $\mathcal{V}_{\lambda,A}$, and $\mathcal{U}_{\lambda,B}$,
$\mathcal{V}_{\lambda,B}$ the irreducible component of $\mathcal{H}%
_{A}^{\otimes n}$ and $\mathcal{H}_{B}^{\otimes n}$ , respectively. Also,
$\mathcal{W}_{\lambda,A}:=\mathcal{U}_{\lambda,A}\otimes\mathcal{V}%
_{\lambda,A}$, $\mathcal{W}_{\lambda,B}:=\mathcal{U}_{\lambda,B}%
\otimes\mathcal{V}_{\lambda,B}$.

Due to \cite{MatsumotoHayashi2}, in terms of \ this decomposition, $\left\vert
\phi\right\rangle ^{\otimes n}$ can be written as%

\begin{equation}
\left\vert \phi\right\rangle ^{\otimes n}=\bigoplus_{\lambda}a_{\lambda
}\left\vert \phi_{\lambda}\right\rangle \left\vert \Phi_{\lambda}\right\rangle
,\label{decomposition}%
\end{equation}
where $\left\vert \phi_{\lambda}\right\rangle \in\mathcal{U}_{\lambda
,A}\otimes\mathcal{U}_{\lambda,B}$, and $\left\vert \Phi_{\lambda
}\right\rangle \in\mathcal{V}_{\lambda,A}\otimes\mathcal{V}_{\lambda,B}$.
While $a_{\lambda}$ and $\left\vert \phi_{\lambda}\right\rangle $ are
dependent on $\left\vert \phi\right\rangle $, $\left\vert \Phi_{\lambda
}\right\rangle $ is a maximally entangled state which does not depend on
$\left\vert \phi\right\rangle $,%
\[
\left\vert \Phi_{\lambda}\right\rangle :=\frac{1}{\sqrt{d_{\lambda}}}%
\sum_{i=1}^{d_{\lambda}}\left\vert f_{i}\right\rangle \left\vert
f_{i}\right\rangle ,
\]
with $\left\{  \left\vert f_{i}\right\rangle \right\}  $ being an orthonormal
complete basis of $\mathcal{V}_{\lambda}$, and $d_{\lambda}:=\dim
\mathcal{V}_{\lambda}$.

Observe that linear span of the state vectors in the of (\ref{decomposition})
is the symmetric subspace of $\left(  \mathcal{H}_{A}\otimes\mathcal{H}%
_{B}\right)  ^{\otimes n}$. Therefore, denoting  the projector on this
subspace by $\Pi^{n}$, we have
\begin{equation}
\Pi^{n}=\bigoplus_{\lambda}\mathcal{U}_{\lambda,A}\otimes\mathcal{U}%
_{\lambda,B}\otimes\left\vert \Phi_{\lambda}\right\rangle \left\langle
\Phi_{\lambda}\right\vert .\label{project-sym}%
\end{equation}

\section{Optimal LOCC of maximally entangled state}

\cite{Tsuda} treats the problem of testing whether the given state $\rho$ is
the $d$dimensional maximally entangled state
\[
\left\vert \Phi\right\rangle :=\frac{1}{\sqrt{d}}\sum_{i=1}^{d}\left\vert
f_{i}\right\rangle \left\vert f_{i}\right\rangle ,
\]
and found out a protocol whose the probability $P_{acc}$ of accepting the
hypothesis equals
\begin{equation}
P_{acc}=\frac{\left\langle \Phi\right\vert \rho\left\vert \Phi\right\rangle
+\frac{1}{\left(  d\right)  ^{2}}}{1+\frac{1}{\left(  d\right)  ^{2}}%
}.\label{tsuda-acc}%
\end{equation}

When $d$ is very large,
\[
P_{acc}\approx\left\langle \Phi\right\vert \rho\left\vert \Phi\right\rangle ,
\]
the RHS of which is the accepting probability of globally optimal one-sided
test. 

\section{Protocol}

Observe the globally optimal test, $\Pi^{n}$, is equivalent to the composition
of the projector $\mathcal{W}_{\lambda,A}\otimes\mathcal{W}_{\lambda,B}$
followed by $\mathbf{I}_{\mathcal{U}_{\lambda,A}\otimes\mathcal{U}_{\lambda
,B}}\otimes\left\vert \Phi_{\lambda}\right\rangle \left\langle \Phi_{\lambda
}\right\vert $. While the former is done by an LOCC, the latter cannot be
implemented by LOCC. Hence, instead, we perform the asymptotically optimal
test of the maximally entangled state in \cite{Tsuda}. So, our protocol is:

\begin{description}
\item[(i)] A and B applies the projective measurement  $\left\{
\mathcal{W}_{\lambda,A}\right\}  _{\lambda}$ and $\left\{  \mathcal{W}%
_{\lambda,B}\right\}  _{\lambda}$, respectively. 

\item[(ii)] Do the test for maximally entangled state to $\mathrm{tr}%
_{\mathcal{U}_{\lambda,A}}\rho_{n,\lambda}$, where  $p_{\lambda}%
:=\mathrm{tr}\,\rho^{\otimes n}\,\mathcal{W}_{\lambda,A}\otimes\mathcal{W}%
_{\lambda,B}$ and $\rho_{n,\lambda}:=\frac{1}{p_{\lambda}}\mathcal{W}%
_{\lambda,A}\otimes\mathcal{W}_{\lambda,B}\,\rho^{\otimes n}\,\mathcal{W}%
_{\lambda,A}\otimes\mathcal{W}_{\lambda,B}$.
\end{description}

\subsection{Peformance of the protocol}

In this subsection, it is proved that our protocol is asymptotically as good
as globally optimal test, $\Pi^{n}$. If the given state is a pure state,
obviously the acceptance probability $P_{opt}^{n}$ of the test $\Pi^{n}$ is 1.
If the input is not a pure state, due to \ref{grep-type-2}, we have
\begin{align*}
-\lim_{n\rightarrow\infty}\frac{1}{n}\log P_{opt}^{n} &  =D\left(  \,\left(
1,0,\cdots,0\right)  \,|\,|\,\boldsymbol{p}\right)  \\
&  =-\log p_{1}.
\end{align*}
Also, by (\ref{project-sym}), when the given state is $\rho^{\otimes n}$,
\begin{align*}
P_{opt}^{n} &  :=\sum_{\lambda}\mathrm{tr}\rho^{\otimes n}\mathcal{U}%
_{\lambda,A}\otimes\mathcal{U}_{\lambda,B}\otimes\left\vert \Phi_{\lambda
}\right\rangle \left\langle \Phi_{\lambda}\right\vert \\
&  =\sum_{\lambda}p_{\lambda}\mathrm{tr}\left\langle \Phi_{\lambda}\right\vert
\rho_{n,\lambda}\left\vert \Phi_{\lambda}\right\rangle .
\end{align*}

Below, we will show our LOCC test is asymptotically equivalent to this
globally optimal test.  On the other hand, due to \ref{tsuda-acc}, our test
will accept the input $\rho_{n}$ with the probability%
\[
P_{\ast}^{n}:=\sum_{\lambda}p_{\lambda}\frac{\mathrm{tr}\left\langle
\Phi_{\lambda}\right\vert \rho_{n,\lambda}\left\vert \Phi_{\lambda
}\right\rangle +\frac{1}{\left(  d_{\lambda}\right)  ^{2}}}{1+\frac{1}{\left(
d_{\lambda}\right)  ^{2}}}.
\]
If the given state $\rho$ is a pure state,
\[
P_{\ast}^{n}=\sum_{\lambda}p_{\lambda}\frac{1+\frac{1}{\left(  d_{\lambda
}\right)  ^{2}}}{1+\frac{1}{\left(  d_{\lambda}\right)  ^{2}}}=1.
\]
Suppose $\rho$ is not a pure state. Observe
\begin{align*}
P_{\ast}^{n} &  \leq\sum_{\lambda}p_{\lambda}\left(  \mathrm{tr}\left\langle
\Phi_{\lambda}\right\vert \rho_{n,\lambda}\left\vert \Phi_{\lambda
}\right\rangle +\frac{1}{\left(  d_{\lambda}\right)  ^{2}}\right)  \\
&  =P_{opt}^{n}+\sum_{\lambda}\frac{p_{\lambda}}{\left(  d_{\lambda}\right)
^{2}},
\end{align*}
where
\begin{align*}
\sum_{\lambda}\frac{p_{\lambda}}{\left(  d_{\lambda}\right)  ^{2}} &
=\sum_{\lambda}\frac{\mathrm{tr}\,\rho^{\otimes n}\,\mathcal{W}_{\lambda
,A}\otimes\mathcal{W}_{\lambda,B}}{\left(  d_{\lambda}\right)  ^{2}}\\
&  \leq\sum_{\lambda}\frac{p_{1}^{n}\left(  \dim\,\mathcal{W}_{\lambda
,A}\right)  ^{2}}{\left(  d_{\lambda}\right)  ^{2}}\\
&  =p_{1}^{n}\sum_{\lambda}\left(  \dim\,\mathcal{U}_{\lambda,A}\right)
^{2}\\
&  =p_{1}^{n}\sum_{\lambda}\left(  \frac{\prod_{i<j}\left(  \lambda
_{i}-\lambda_{j}-i+j\right)  }{\prod_{i=1}^{d-1}\left(  d-i\right)  !}\right)
^{2}\\
&  \leq p_{1}^{n}\left(  n+1\right)  ^{d}n^{d^{2}}.
\end{align*}
(Also, one may use the relation
\begin{align*}
&  \sum_{\lambda}\left(  \dim\,\mathcal{U}_{\lambda,A}\right)  ^{2}\\
&  =\dim\left(  \text{symmetric subspace of }\left(
\mathbb{C}
^{d^{2}}\right)  ^{\otimes n}\right)  \\
&  \leq\left(  n+1\right)  ^{d^{2}}%
\end{align*}
)

Therefore, even if the state $\rho$ is not a pure state,
\[
-\lim_{n\rightarrow\infty}\frac{1}{n}\log P_{\ast}^{n}\geq-\lim_{n\rightarrow
\infty}\frac{1}{n}\log P_{acc}^{n}.
\]
Since the other side of inequality is trivial, we have
\[
-\lim_{n\rightarrow\infty}\frac{1}{n}\log P_{\ast}^{n}=-\lim_{n\rightarrow
\infty}\frac{1}{n}\log P_{acc}^{n}.
\]

Therefore, regardless $\rho$ is pure or not, our LOCC protocol very closely
approximates the globally optimal protocol when $n$ is large.

\appendix

\section{Group representation theory}

\label{appendixA}

\begin{lemma}
\label{lem:decohere} Let $U_{g}$ and $U_{g}^{\prime}$ be an irreducible
representation of $G$ on the finite-dimensional space $\mathcal{H}$ and
$\mathcal{H}^{\prime}$, respectively. We further assume that $U_{g}$ and
$U_{g}^{\prime}$ are not equivalent. If a linear operator $A$ in
$\mathcal{H}\oplus\mathcal{H}^{\prime}$ is invariant by the transform
$A\rightarrow U_{g}\oplus U_{g}^{\prime}AU_{g}^{\ast}\oplus U_{g}^{^{\prime
}\ast}$ for any $g$, $\mathcal{H}A\mathcal{H^{\prime}}=0$ ~\cite{GW}.
\end{lemma}

\begin{lemma}
\label{lem:shur} (Shur's lemma~\cite{GW}) Let $U_{g}$ be as defined in
lemma~\ref{lem:decohere}. If a linear map $A$ in $\mathcal{H}$ is invariant by
the transform $A\rightarrow U_{g}AU_{g}^{\ast}$ for any $g$, $A=c\mathrm{Id}%
_{\mathcal{H}}$.
\end{lemma}

\section{Representation of symmetric group and SU}

Due to \cite{GW}, we have%

\begin{align}
\dim\mathcal{U}_{\lambda} &  =\frac{\prod_{i<j}\left(  l_{i}-l_{j}\right)
}{\prod_{i=1}^{d-1}\left(  d-i\right)  !},\label{dim-representation-1}\\
d_{\lambda} &  =\dim\mathcal{V}_{\lambda}=\frac{n!}{\prod_{i=1}^{d}\left(
\lambda_{i}+d-i\right)  !}\prod_{i<j}\left(  l_{i}-l_{j}\right)
,\label{dim-representation-2}%
\end{align}
with $l_{i}:=\lambda_{i}+d-i$. \ It is easy to show%
\begin{equation}
\log\dim\mathcal{U}_{\lambda}\leq d^{2}\log n.\label{dim-zero-rate}%
\end{equation}

Below,
\[
\left\vert \phi\right\rangle =\sum_{i=1}^{d}\sqrt{p_{i}}\left\vert
e_{i}\right\rangle \left\vert e_{i}\right\rangle ,
\]
where $\left\{  \left\vert e_{i}\right\rangle \right\}  _{i}$ is an
orthonormal basis of $\mathcal{H}$. With  $a_{\lambda}^{\phi}=\mathrm{Tr}%
\left\{  \mathcal{W}_{\lambda,A}\left(  \mathrm{Tr}_{B}|\phi\rangle\langle
\phi|\right)  ^{\otimes n}\right\}  $,%

\begin{align}
\left\vert \frac{\log d_{\lambda}}{n}-\mathrm{H}\left(  \frac{\lambda}%
{n}\right)  \right\vert  &  \leq\frac{d^{2}+2d}{2n}\log
(n+d),\label{grep-type-1}\\
\sum_{\frac{\lambda}{n}\in\mathrm{R}}a_{\lambda}^{\phi} &  \leq\left(
n+1\right)  ^{d\left(  d+1\right)  /2}\exp\left\{  -n\min_{\boldsymbol{q}%
\,\in\mathrm{R}}\mathrm{D}\left(  \boldsymbol{q}||\boldsymbol{p}\right)
\right\}  ,\label{grep-type-2}%
\end{align}
where $\mathrm{R}$ is an arbitrary closed subset \cite{MatsumotoHayashi}.


\begin{thebibliography}{9}                                                                                                %
\bibitem {GW}R.~Goodman and N.~Wallach, \textit{Representations and Invariants
of the Classical Groups}, (Cambridge University Press, 1998.

\bibitem {Tsuda}M. Hayashi, K. Matsumoto, and Y. Tsuda, "A study of
LOCC-detection of a maximally entangled state using hypothesis testing", J. of
Phys. A, 39 14427-14446 (2006).

\bibitem {MatsumotoHayashi}M.~Hayashi and K.\thinspace Matsumoto,
\textquotedblleft Quantum universal variable-length source
coding,\textquotedblright\ Phys. Rev. A 66, 022311(2002).

\bibitem {MatsumotoHayashi2}K. Matsumoto and M. Hayashi, "Universal
distortion-free entanglement concentration", Phys. Rev. A 75, 062338 (2007).
\end{thebibliography}
\end{document}